\documentclass[aps,prd,groupedaddress,showpacs,graphicx,nofootinbib]{revtex4}
\usepackage{amssymb,graphics,graphicx,color}

\begin{document}

\title{Distinguishing between Extra Natural Inflation and Natural Inflation after BICEP2}

\author{Kazunori Kohri$^{1}$}\email{kohri@post.kek.jp}
\author{C. S. Lim$^{2}$}\email{lim@lab.twcu.ac.jp}
\author{Chia-Min Lin$^{3}$}\email{lin@chuo-u.ac.jp}

\affiliation{$^1$Cosmophysics Group, Theory Center, IPNS
KEK, and The Graduate University for Advanced Studies (Sokendai), 1-1
Oho, Tsukuba 305-0801, Japan}
\affiliation{$^2$Department of Mathematics, Tokyo Woman's Christian University, Tokyo 167-8585, Japan}
\affiliation{$^3$Department of Physics, Chuo University, Bunkyo-ku, Tokyo 112, Japan}

\date{Draft \today}

\begin{abstract}
In this paper, we carefully calculated the tensor-to-scalar ratio, the running spectral index, and the running of running spectrum for (extra) natural inflation in order to compare with recent BICEP2 data, PLANCK satellite data and future 21 cm data. We discovered that the prediction for running spectral index and the running of running spectrum in natural inflation is different from that in the case of extra natural inflation. Near future observation for the running spectral index can only provide marginal accuracy which may not allow us distinguishing between extra natural inflation from natural inflation clearly unless the experimental accuracy can be further improved.


\end{abstract}
\maketitle

\section{Introduction}

\large
\baselineskip 18pt

Inflation \cite{Starobinsky:1980te, Sato:1980yn, Guth:1980zm, Linde:1981mu, Albrecht:1982wi} (see \cite{Lyth:2009zz} for a review) is becoming a standard model for the very early universe. Recent report from BICEP2 gives a tensor-to-scalar ratio (at 1-$\sigma$ level) to be \cite{Ade:2014xna}
\begin{equation}
r=0.20^{+0.07}_{-0.05}
\end{equation}
which seems to prefer a large field inflation model (which means the field value during inflation is larger than Planck scale). One interesting realization of a large field inflation model is natural inflation \cite{Freese:1990rb} where the inflaton field is a pseudo-Nambu-Goldstone boson (PNGB). Recently a comparison of natural inflation with BICEP2 is made in \cite{Freese:2014nla, Czerny:2014qqa, Harigaya:2014eta, Choi:2014rja, Higaki:2014pja, Kappl:2014lra, Bachlechner:2014hsa, Long:2014dta}. There is also an extra natural inflation \cite{ArkaniHamed:2003wu} (see also \cite{Furuuchi:2013ila, Inami:2009bs} for some recent model building) which is usually regarded as another realization of natural inflation via extra dimensional phenomenon. Those models (both natural and extra natural inflation) can be distinguished from chaotic inflation \cite{Creminelli:2014oaa, Chiba:2014gfa}. 
In this paper, we point out that the prediction for the running spectral index $\alpha$ and the running of running spectrum $\beta$ are different between natural inflation and extra natural inflation. The observation of those parameters will be improved by future 21 cm data \cite{Kohri:2013mxa}. However, the expected accuracy is marginal and we have to wait for further improvement for the data in order to achieve decisive conclusion.


\section{Scale dependence of the spectrum}
\label{sec1}
The spectrum generated during inflation is given by the following standard formula (for simplicity, we set the reduced Planck mass $M_P=1$),
\begin{equation}
P=\frac{1}{12\pi^2}\frac{V^3}{(V^\prime)^2},
\label{eq1}
\end{equation}
where $V$ is the inflaton potential and the derivative is with respect to the inflaton field $\phi$. The simplest way to characterize the scale dependence of the spectrum is to investigate the spectral index $n$, defined by
\begin{equation}
n -1 \equiv \frac{d \ln P}{d \ln k}.
\label{eq2}
\end{equation} 
By using the handy relation,
\begin{equation}
\frac{d}{d \ln k} = -\frac{V^\prime}{V}\frac{d}{d \phi},
\label{eq3}
\end{equation}
it is easy to show from Eqs.~(\ref{eq1}) and (\ref{eq2}) that
\begin{equation}
n=1+2\eta-6\epsilon,
\end{equation}
where the slow-roll parameters are defined as 
\begin{equation}
\epsilon \equiv \frac{1}{2}\left(\frac{V^\prime}{V}\right)^2, \;\;\;\; \eta \equiv \frac{V^{\prime\prime}}{V}.
\end{equation}
We can define higher order slow-roll parameters as
\begin{equation}
\xi \equiv \frac{V^\prime V^{\prime\prime\prime}}{V^2}, \;\;\;\; \sigma  \equiv \frac{(V^\prime)^2 V^{\prime\prime\prime\prime}}{V^3}, \;\;\;\; \chi \equiv \frac{(V^\prime)^2 V^{\prime\prime\prime\prime\prime}}{V^4}.
\end{equation}
By using Eq.~(\ref{eq3}), the running spectrum is given by
\begin{equation}
\alpha \equiv \frac{d n}{d \ln k} = 16 \epsilon \eta -24 \epsilon^2 - 2\xi.
\end{equation}
The running of running spectrum can also be obtained as
\begin{equation}
\beta \equiv \frac{d \alpha}{d \ln k} = 2\eta \xi + 2 \sigma -24 \epsilon \xi -32 \epsilon \eta^2 +192 \epsilon ^2 \eta -192 \epsilon^3.
\end{equation}
Usually we have the hierarchy 
\begin{equation}
|\sigma| \ll |\xi| \ll |\eta|.
\end{equation}
This is also the case for (extra) natural inflation.

According to Planck data \cite{Ade:2013ydc}, $\alpha=-0.0134 \pm 0.0090$ (at 1-$\sigma$, i.e., $-0.0134 \pm 0.0180$ at 2-$\sigma$) and $-0.03<\beta<+0.06$ at 2-$\sigma$.  Near future measurement of 21 cm fluctuations will improve the accuracy by one or two orders \cite{Kohri:2013mxa}.
Ideally, we would have
\begin{equation}
\Delta \alpha \sim \Delta \beta \sim {\cal O}(10^{-4}),
\label{future}
\end{equation}
at 2$\sigma$.

\section{natural inflation and extra natural inflation}
\label{IR}
Natural inflation assumes the inflaton field is a pseudo Nambu Goldstone Boson (PNGB), with the inflaton potential given by
\begin{equation}
V(\phi)=V_0\left(1-\cos\left(\frac{\phi}{f}\right)\right),
\label{natural}
\end{equation}
where $f$ is the spontaneous symmetry breaking scale. In order for inflation to occur, we need $f>M_P$ which is outside the range of validity of an effective field theory description. This motivates the introduction of extra natural inflation (although other ways to evade the problem exist, such as assisted inflation \cite{Liddle:1998jc}, N-flation \cite{Dimopoulos:2005ac}, the Kim-Nilles-Peloso mechanism \cite{Kim:2004rp} and the axion monodromy \cite{Silverstein:2008sg}). The idea of extra natural inflation is that a similar form of the potential can be obtained in the framework of a 5d model with the extra dimension compactified on a circle of radius $R$. The extra component $A_5$ of an abelian gauge field plays the role of the inflaton and in this case the effective decay constant is given by,
\begin{equation}
f= \frac{1}{2\pi g_{4d}R}=\frac{1}{\sqrt{2\pi R}g_{5d}},
\end{equation}
where $g_{5d}$ is the 5D gauge coupling and $g_{4d}^2=g_{5d}^2/(2 \pi R)$ is the 4D gauge coupling. Note that the condition $f>M_P$ can be achieved by a sufficiently small $g_{4d}$ for a $R$ larger than Planck length.

In the case of extra natural inflation, the potential is not really given by Eq.~(\ref{natural}), but we should replace it as
\begin{equation}
V=V_0 \left( 1-\sum_{n=1}^{\infty} \frac{\cos (n\frac{\phi}{f})}{n^5} \right).
\label{extranatural}
\end{equation}
The potential in Eq.~(\ref{natural}) is obtained if we approximate it by choosing only $n=1$ and neglecting other terms which are suppressed by the factor $1/n^5$. This is reasonable if we are calculating slow-roll parameters $\epsilon$ and $\eta$. However, those higher order terms becomes important when we are calculating higher order slow-roll parameters $\sigma$ and $\chi$. 

We now argue why the contributions from larger $n$ become important when we consider higher order slow-roll parameters. We first briefly see the origin of the term $- V_0 \sum_{n=1}^{\infty} \frac{\cos (n\frac{\phi}{f})}{n^5}$ in Eq.~(\ref{extranatural}). It is nothing but the effective potential in one of the proposals of physics beyond the standard model, ``gauge-Higgs unification", where Higgs is identified with the extra-space component of higher-dimensional gauge field $A_5$ \cite{Manton:1979kb, Hosotani:1983xw, Hosotani:1983vn, Hatanaka:1998yp, Kubo:2001zc}. In the simplest case of 5-dimensional space-time, the Higgs is not allowed to have potential at the classical level, just because gauge field itself is gauge-variant quantity and its polynomials are forbidden by the local gauge symmetry of the theory. At the quantum level, however, concrete calculation shows that the potential such as the one mentioned above, $- V_0 \sum_{n=1}^{\infty} \frac{\cos (n\frac{\phi}{f})}{n^5}$, is induced, where $f \sim 1/R$ with $R$ being the size of the extra space, and we can even discuss the spontaneous symmetry breaking due to the VEV of that Higgs field. To be strict, the potential is due to the quantum effects of particles with zero bulk mass (higher dimensional mass). 

At the first glance, this situation is a little strange, since the VEV, i.e. a constant gauge field  (recall that Higgs is originally a gauge field) is equivalent to vanishing ``electro-magnetic" fields (though the gauge symmetry needs not to be U(1)$_{em}$) and therefore is just a ``pure gauge" configuration. As the matter of fact, the constant field has physical meaning as a sort of Aharonov-Bohm (AB) phase or Wilson-loop phase when the extra space is non-simply connected space, such as the circle $S^{1}$. Namely, it may be interpreted as the measure of ``magnetic" flux penetrating inside the circle. 

As far as the Higgs field is understood as a AB-phase, every observables are expected to be periodic in the field. That is why  trigonometric functions appear in the effective potential (\ref{extranatural}). In fact, we realize 
\begin{equation}
\cos \left(n\frac{\phi}{f}\right) =\mbox{Re}\{ (e^{i\frac{\phi}{f}})^{n} \},  
\end{equation}
where $n = 1,2, \ldots$ denotes how many times the loop of the Feynman diagram is wrapped along the circle $S^{1}$. 
Namely, $n$ is ``winding number" rather than the integer to denote Kaluza-Klein modes, which appears after the technique of ``Poisson resummation" is used. Being wrapped $n$ times, the particles propagating along the loop will ``feel" AB phase $n$ times. 

Now we realize some peculiar features of the potential (\ref{extranatural}). If we perform Taylor-expansion of the potential at the origin $\phi = 0$, the coefficients of the higher order terms are all divergent. For instance, 
\begin{equation}
\frac{d^{4}V}{d\phi^{4}}|_{\phi = 0} = - \frac{V_0}{f^{4}} \sum_{n=1}^{\infty} \frac{1}{n},   
\end{equation}
which is clearly a divergent sum. The divergence appears at the fourth-derivative for the first time and becomes more serious as 
the order of the derivative increases. Such feature is not shared by natural inflation scenario and is specific to extra natural inflation.  

The physical interpretation of the origin of the divergence is rather simple; it is a sort of infra-red (IR) divergence coming from the fact that the particle contributing to the quantum effect is massless in the higher-dimensional sense. In fact, the fourth derivative discussed above corresponds to the 4-point function of Higgs field, which is obtained from a Feynman diagram with 4 propagators of massless fermion, e.g., and the integral over the ``loop-momentum" $k$ gives divergence at the vicinity of small 4-momentum for the Kaluza-Klein zero mode: 
\begin{equation}
\int_{0}^{\Lambda} \frac{k^{3} dk}{k^{4}} =  \infty,   
\end{equation}
where $\Lambda$ is UV-cutoff, though UV-divergence eventually does not appear in the gauge-Higgs unification after the the contributions from all non-zero Kaluza-Klein modes are summed up \cite{Hatanaka:1998yp}.          

Now, it is easily understood that once we add bulk masses to the particles contributing to the quantum correction, such IR divergences can be avoided. We note that the gauge fields are not allowed to have bulk masses, though. 

Actually, such IR divergences turn out not to appear even for the case of vanishing bulk mass, once we avoid the origin, 
namely if we perform Taylor-expansion around some non-vanishing value of the field $\phi$. This is because the non-vanishing field $\phi$ itself behaves as VEV of the Higgs, providing non-vanishing 4-dimensional mass for the particles inside the loop. As a typical example the fourth derivative now behaves as 
\begin{equation}
\frac{d^{4}V}{d\phi^{4}} = - \frac{V_0}{f^{4}} \sum_{n=1}^{\infty} \frac{\cos (n\frac{\phi}{f})}{n} 
= \frac{V_0}{f^{4}} \log \left[ 2 \sin \left( \frac{\phi}{2f} \right) \right], 
\label{fourth}  
\end{equation}
which becomes logarithmically divergent as $\phi \to 0$, which is nothing but the IR singularity we discussed above. In (\ref{fourth}), a formula 
\begin{equation}
\sum_{n=1}^{\infty} \frac{\cos (nx)}{n} 
= - \log \left[2 \sin \left(\frac{x}{2}\right)\right] 
\label{fourier}
\end{equation}
has been used \cite{Iwanami}. 

We easily find that all derivatives of the potential are IR-finite, since higher derivatives obtained by taking derivatives of (\ref{fourth}) successively should be all finite for $\phi \neq 0$. We thus conclude that for non-vanishing $\phi$, relevant for the inflation scenario, all higher derivatives are obtained as finite values. We at the same time point out that when $\phi \ll f$ there appears a tendency that higher derivatives become significant.

\section{Inflation Analysis}
\label{analysis}

In the following, we will analyze (extra) natural inflation in detail. 
To our knowledge, this detailed analytical calculation for extra natural inflation is done for the first time.
We use Eq.~(\ref{natural}) to calculate $V^\prime$ and $V^{\prime\prime}$ since the higher order terms $n>1$ in Eq.~(\ref{extranatural}) are suppressed. However, for $V^{\prime\prime\prime}$ and higher derivative terms we shall use Eq.~(\ref{extranatural}) for extra natural inflation. It is interesting to find that this strategy allows us to carry out the calculation in a (rather elegant) analytical way.
The slow roll parameters are given by
\begin{equation}
\epsilon=\frac{1}{2f^2}\frac{1+\cos(\frac{\phi}{f})}{1-\cos(\frac{\phi}{f})}, \;\;\;\;\; \eta=\frac{1}{2f^2}\frac{\cos(\frac{\phi}{f})+\cos(\frac{\phi}{f})}{1-\cos(\frac{\phi}{f})}.
\end{equation}
It is therefore clear that $\epsilon$ is larger than $\eta$ and inflation ends when $\epsilon(\phi_e)=1$, where $\phi_e$ denotes the field value at the end of inflation. This gives
\begin{equation}
\cos\left(\frac{\phi_e}{f}\right)=\frac{1-\frac{1}{2f^2}}{1+\frac{1}{2f^2}}.
\end{equation}
The spectral index is given by
\begin{equation}
n=1+2\eta-6\epsilon=1-\frac{1}{f^2}\frac{3+\cos(\frac{\phi}{f})}{1-\cos(\frac{\phi}{f})}.
\end{equation}
The number of e-folds is 
\begin{equation}
N=\int_{\phi_e}^{\phi} \frac{V}{V^\prime}d\phi=2f^2 \ln \frac{\cos(\frac{\phi_e}{2f})}{\cos(\frac{\phi}{2f})}.
\end{equation}
Therefore 
\begin{equation}
\cos\left(\frac{\phi}{f}\right)=2\frac{1}{e^{\frac{N}{f^2}}(1+\frac{1}{2f^2})}-1,
\end{equation}
This allows us to calculate $n$ and $\epsilon$ as a function of $f$ when we fix $N$ and we will use $N=60$.

For natural inflation, we have
\begin{equation}
V^{\prime\prime\prime}=-V_0\frac{1}{f^3}\sin\left(\frac{\phi}{f}\right).
\end{equation}
On the other hand, for extra natural inflation, we have
\begin{equation}
V^{\prime\prime\prime}=-V_0\frac{1}{f^3}\sum_{n=1}^{\infty}\frac{\sin(n\frac{\phi}{f})}{n^2} \sim V_0 \frac{1}{f^3}\left(\frac{\phi}{f}\ln \left(\frac{\phi}{f}\right)-\frac{\phi}{f}\right),
\end{equation}
where we have used the Fourier series \cite{Iwanami}
\begin{equation}
\sum_{n=1}^{\infty}\frac{\sin(nx)}{n^2}=-x\ln(x)+x+\frac{x^3}{18 \cdot 2^2}+\frac{x^5}{900\cdot 2^4}+\frac{x^7}{19845\cdot 2^6}+\cdots.
\end{equation}
We keep only the first two terms because $\frac{\phi}{f}<\pi$. We can hence calculate the higher order slow roll parameter $\xi$. For natural inflation, we have
\begin{equation}
\xi=-\frac{2}{f^2}\epsilon.
\end{equation} 
For extra natural inflation, we have
\begin{equation}
\xi=\frac{1}{f^4}\frac{\sin(\frac{\phi}{f})(\frac{\phi}{f}\ln(\frac{\phi}{f})-\frac{\phi}{f})}{(1-\cos(\frac{\phi}{f}))^2}.
\end{equation}
Finally, for natural inflation, we have
\begin{equation}
V^{\prime\prime\prime\prime}=-V_0\frac{1}{f^4}\cos\left(\frac{\phi}{f}\right).
\end{equation}
For extra natural inflation,
\begin{equation}
V^{\prime\prime\prime\prime}=-V_0\frac{1}{f^4}\sum_{n=1}^{\infty}\frac{\cos(n\frac{\phi}{f})}{n}=V_0\frac{1}{f^4}\ln \left[2\sin \left( \frac{\phi}{2f}\right)\right],
\end{equation}
where we have used Eq.~(\ref{fourier}).
This allows us to calculate the higher order slow roll parameter $\sigma$ as
\begin{equation}
\sigma=\frac{1}{f^4}\epsilon \frac{\ln(2-2\cos(\frac{\phi}{f}))}{1-\cos(\frac{\phi}{f})}.
\end{equation}

\begin{figure}[t]
  \centering
\includegraphics[width=0.6\textwidth]{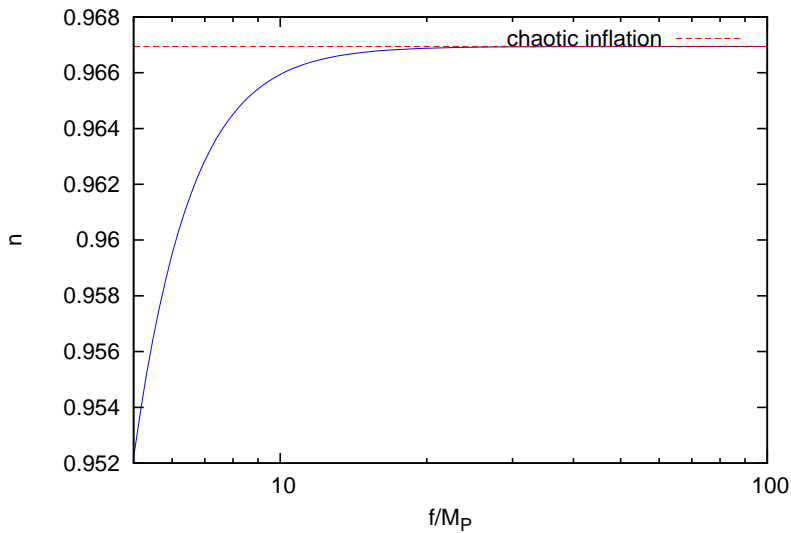}
  \caption{The spectral index $n$ as a function of $f$. Note that $n$ becomes the same as (quadratic) chaotic inflation for large $f/M_P$.}
  \label{fig1}
\end{figure}

\begin{figure}[t]
  \centering
\includegraphics[width=0.6\textwidth]{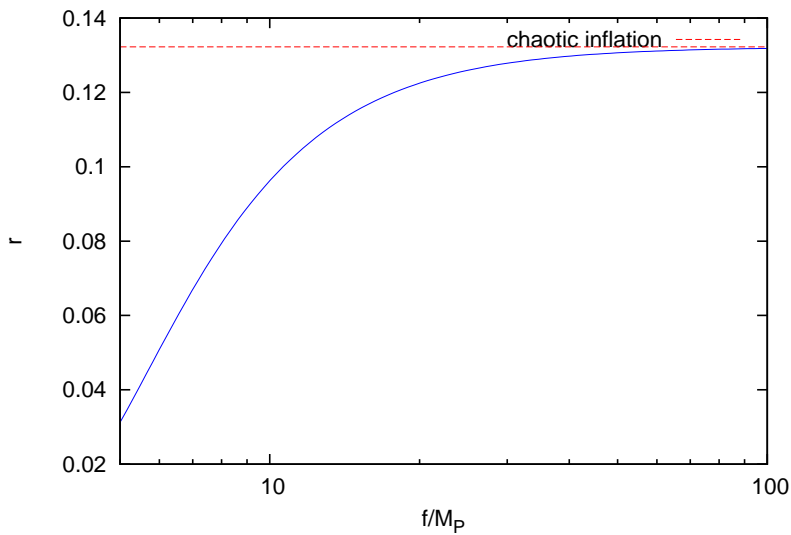}
  \caption{Tensor-to-scalar ratio $r$ as a function of $f$. Note that $r$ becomes the same as chaotic inflation for large $f/M_P$. This result can be compared with BICEP2 and future measurement of primordial gravity waves.}
  \label{fig2}
\end{figure}

\begin{figure}[t]
  \centering
\includegraphics[width=0.6\textwidth]{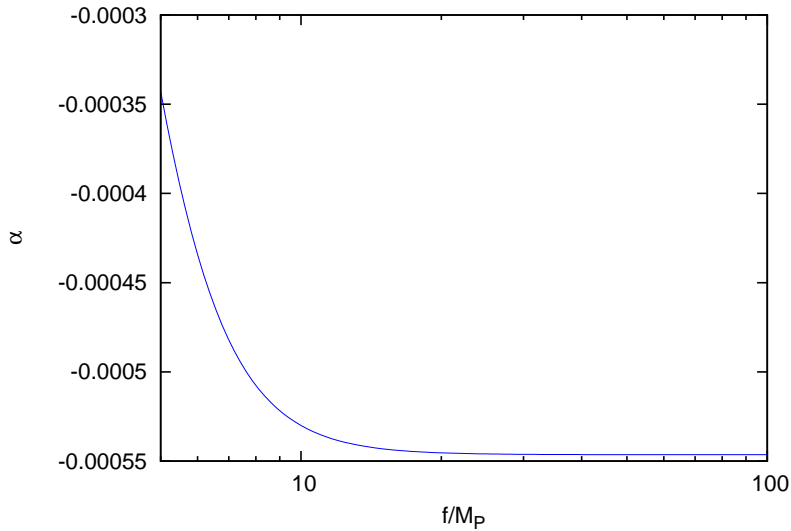}
  \caption{The running spectral index $\alpha$ as a function of $f$ for natural inflation}
  \label{fig3}
\end{figure}

\begin{figure}[t]
  \centering
\includegraphics[width=0.6\textwidth]{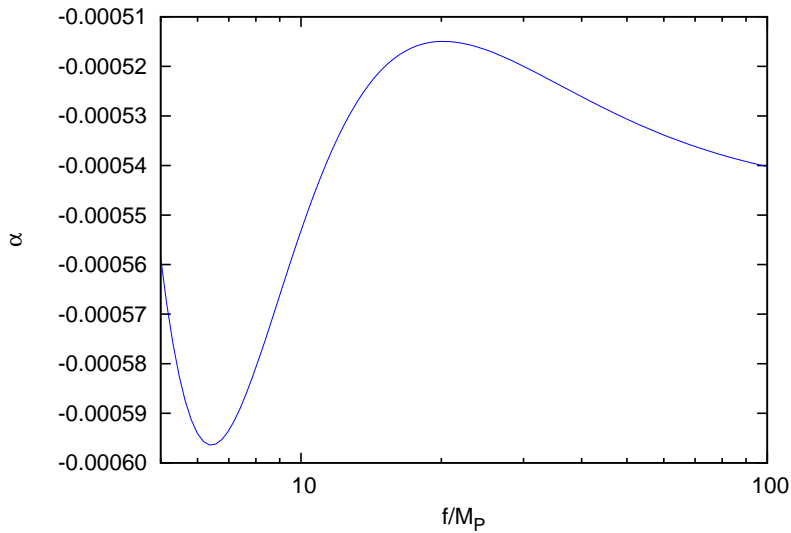}
  \caption{The running spectral index $\alpha$ as a function of $f$ for extra natural inflation}
  \label{fig4}
\end{figure}

\begin{figure}[t]
  \centering
\includegraphics[width=0.6\textwidth]{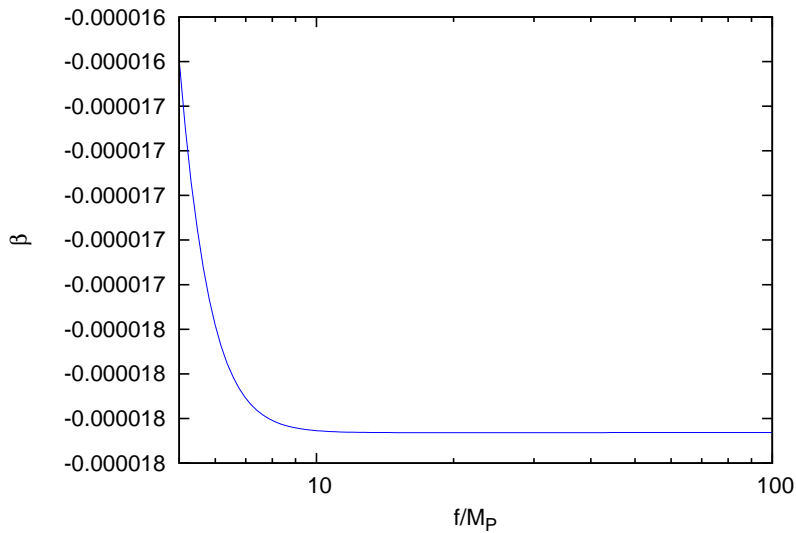}
  \caption{The running of running spectrum $\beta$ as a function of $f$ for natural inflation.}
  \label{fig5}
\end{figure}

\begin{figure}[t]
  \centering
\includegraphics[width=0.6\textwidth]{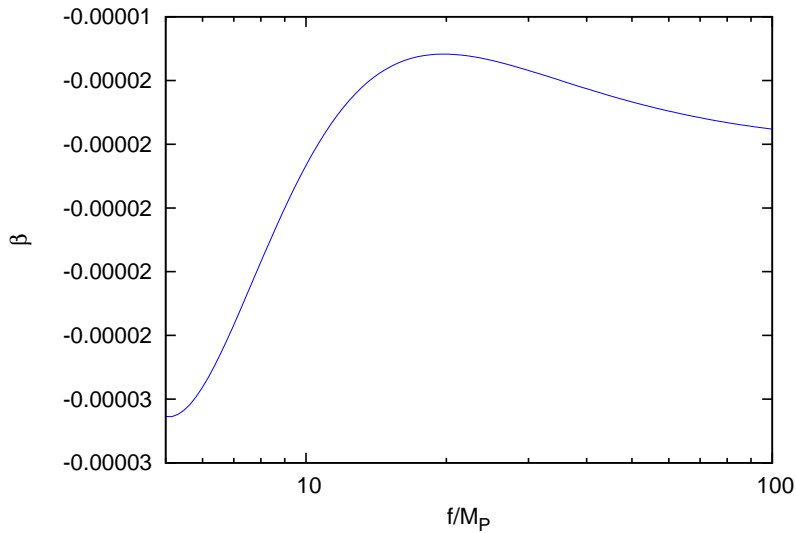}
  \caption{The running of running spectrum $\beta$ as a function of $f$ for extra natural inflation.}
  \label{fig6}
\end{figure}

We can now make plots to show our results. Firstly, the spectral index $n$ is shown in FIG.~\ref{fig1}. The plot shows we should consider $f \gtrsim 5M_P$ and for large $f$, the spectral index approaches $n=0.967$ which coincides with quadratic chaotic inflation.
We plot the tensor-to-scalar ratio $r=16\epsilon$ in FIG.~\ref{fig2}. This shows large $f$ corresponds to higher inflation scale. This result can be compared with experiments such as BICEP2.
The running spectral index $\alpha$ for natural inflation is plotted in FIG.~\ref{fig3} and $\alpha$ for extra natural inflation is plotted in FIG.~\ref{fig4}. From Eq.~(\ref{future}), we can see that it may be possible to distinguish between extra natural inflation and natural inflation if we have $\Delta \alpha \sim 10^{-4}$. However, a conclusive result still have to be confirmed if experimental accuracy can be improved further.
The running of running spectrum $\beta$ for natural inflation is plotted in FIG.~\ref{fig5} and that for extra natural inflation is plotted in FIG.~\ref{fig6}. As we can see from the plots, in order to distinguish those two cases, we need at least $\Delta \beta \sim 10^{-5}$ which may not be available in the near future. On the other hand, it is interesting to note that the predicted $\beta$ for extra natural inflation is not too large to be inconsistent with current experimental data, albeit the "IR divergences" of the fourth derivative near the bottom of the potential described in section \ref{IR}.   
\section{Conclusion and Discussion}
\label{con}

Due to recent report of BICEP2, there is a trend to revive large field inflation models. It is thus an important task to find methods (if there is any) to distinguish among those models.

In this paper, we present an analysis for (extra) natural inflation. We found the prediction of those two models for the running spectral index $\alpha$ and the running of running spectrum $\beta$ are different.
Future experiments of 21 cm fluctuation can give precise measurements of $\alpha$ and $\beta$. In particular the running spectral index $\alpha$ will provide us a marginal accuracy for distinguishing between extra natural inflation and natural inflation. Further improvement of accuracy is needed to obtain a conclusive result.

\section*{Acknowledgement}
This work is partially supported by the Grant-in-Aid for Scientific
research from the Ministry of Education, Science, Sports, and Culture,
Japan, Nos. 21111006, 22244030, 23540327, 26105520, 26247042 (K.K.), 23654090,
23104009 (C.S.L), and 21244036 (C.S.L and C.M.L.), and  by  the Center
for the Promotion of Integrated Science (CPIS) of Sokendai
(1HB5804100) (K.K.).

\end{document}